\documentclass[preprint,showpacs,preprintnumbers,amsmath,amssymb]{revtex4}

\usepackage{graphicx}% Include figure files
\usepackage{dcolumn}% Align table columns on decimal point
\usepackage{bm}% bold math

\begin{document}

\preprint{}

\title{Dimension-Independent Positive-Partial-Transpose Probability Ratios}

\author{Paul B. Slater}% 
\email{slater@kitp.ucsb.edu}
\affiliation{%
ISBER, University of California, Santa Barbara, CA 93106\\
}%
\date{\today}% It is always \today, today,
             %  but any date may be explicitly specified

\begin{abstract}
We conduct quasi-Monte Carlo numerical integrations in two very
high (80 and 79)-dimensional domains --- the parameter spaces of  
rank-9 and rank-8 
{\it qutrit-qutrit} ($9 \times 9$) density matrices. We, then, estimate 
the {\it ratio} of the probability --- in terms of the Hilbert-Schmidt metric --- that a generic rank-9 density matrix has a {\it positive partial transpose} 
(PPT) to the probability that a generic rank-8 density matrix has a PPT 
(a precondition to {\it separability/nonentanglement}).
Close examination of the numerical results generated --- despite certain large
fluctuations --- indicates that the true ratio may, in fact, be 2.
Our earlier investigation 
(eprint quant-ph/0410238) also yielded estimates close to 2 
 of the comparable ratios for {\it qubit-qubit} and
{\it qubit-qutrit} pairs 
(the only two cases where the PPT condition 
{\it fully} implies separability). Therefore, it merits conjecturing
(as \.Zyczkowski was the first to do)
 that such Hilbert-Schmidt (rank-$NM$/rank-$(NM-1)$) 
PPT probability ratios are 
2 for {\it all} $NM$-dimensional quantum systems.
\end{abstract}

% PACS, the Physics and Astronomy
                             % Classification Scheme.
%\keywords{Suggested keywords}%Use showkeys class option if keyword
                              %display desired
\maketitle
Recent numerical analyses of ours  \cite{slaterPRA} 
have raised the possibility  that there exists a certain quantum
information-theoretic relation, {\it independent} of dimension ($NM$), expressible
in the form
\begin{equation} \label{firstdefinition}
\Omega^{HS}_{NM} \equiv 
\frac{P^{[HS,rank-NM]}_{NM}}{P^{[HS,rank-(NM-1)]}_{NM}} =2.
\end{equation}
Here, HS denotes  the {\it Hilbert-Schmidt} (HS) 
metric \cite{ozawa,zs1}. $\Omega^{HS}_{NM}$ is the ratio of the  
probability ($P^{[HS,rank-NM]}_{NM}$) 
that a {\it full} rank $N M \times N M$ density matrix ($\rho$) 
has a {\it 
positive partial transpose}
(PPT) to the probability ($P^{[HS,rank-(NM-1)]}_{NM}$) 
of a density matrix of the same dimension but of
rank one less $(N M -1$) than full rank ($N M$), having a PPT.

The partial transposition (PT)  operation 
takes the $N \times N$ blocks ($M^2$ 
in number) --- or the $M \times M$ blocks ($N^2$ in number) --- of 
$\rho$, and transposes them in 
place \cite[eq. (21)]{horodecki}. If none of the eigenvalues of the 
result is negative, as is necessarily the case with (the original, untransposed) $\rho$ itself, by the properties of a density matrix,
then the partial transposition is said
to be {\it positive}. 
The Hilbert-Schmidt distance between two density matrices ($\rho_{1}$ 
and $\rho_{2}$) is defined as the Hilbert-Schmidt (Frobenius) norm of their
difference \cite[eq. (2.3)]{zs1},
\begin{equation}
D_{HS}(\rho_{1},\rho_{2}) =||\rho_{1}-\rho_{2}|| =\sqrt{\mbox{Tr} |(\rho_{1}-\rho_{2})^2|}.
\end{equation}

The probabilities in the definition (\ref{firstdefinition}) have a {\it geometric} interpretation. The numerator ($P^{[HS,rank-NM]}_{NM}$) 
is itself the ratio of the Hilbert-Schmidt 
 $(N^2 M^2-1)$-dimensional volume of PPT
states to the volume of {\it all} (PPT {\it and} non-PPT) states. 
The denominator ($P^{[HS,rank-(NM-1)]}_{NM}$) is the ratio of the $(N^2 M^2 -2)$-dimensional ``hyperarea'' (bounding the volume) of rank-$(N M-1)$ PPT states to the hyperarea occupied by 
all rank-$(N M-1)$ states. (Exact formulas have recently become
available for these {\it total} HS volumes and hyperareas \cite{zs1} 
(cf. \cite{hans1}), but {\it not} obviously
for their PPT subsets --- otherwise there would be no need for
our {\it numerical} investigations here, as the true value of 
(\ref{firstdefinition}) could, then, be simply directly computed.)
($\Omega^{HS}_{NM}$ can also be seen as the ratio of two hyperarea-to-volume
ratios, one ratio 
based on all the states, and the other just on the PPT states.)

We specifically studied in \cite{slaterPRA} 
the cases $N=2,M=2$ (qubit-qubit pairs) and $N=2,M=3$ (qubit-qutrit pairs). 
Only in those two specific low-dimensional cases is having
a PPT 
{\it fully} equivalent 
(by the Peres-Horodecki criterion \cite{peres,horodecki}) to the
property of {\it separability} (non-entanglement). 
For $N M > 6$, the PPT condition is only necessary, 
but not sufficient for separability, and as Aubrun and Szarek have recently established becomes ``weaker and weaker [in detecting separability] 
as the dimension increases'' \cite{szarek}.
(Of course, it would be of interest as well 
to study for $NM>6$,  ratios based on
separability probabilities, rather than  PPT probabilities, but that seems more
difficult still (cf. \cite{zapatrin,doherty}).) A state which has a PPT, and yet is not separable, exhibits ``bound entanglement'' \cite{terhal} 
(cf. \cite{terhal2}).

Presented initially with only our qubit-qutrit analyses, 
\.Zyczkowski proceeded to
theorize --- without yet a formal proof, however --- that
$\Omega^{HS}_{NM}$ is {\it always}
 equal to 2 for any $N=M$, and even possibly for
$N \neq M$. (Perhaps we might also observe that when 
these conjectures were brought to the attention of S. Szarek,
he commented that ``the relationship can not
be too hard [to prove] if true''.)

To be more specific, we had reported in 
\cite[sec.~VI C 2]{slaterPRA} --- based on a quasi-Monte Carlo (Tezuka-Faure 
\cite{faure,giray1}) 
numerical integration procedure, 
employing  $4 \times 10^8$  
sample points in 15- and 14-dimensional spaces (unit hypercubes) --- an estimate in the {\it qubit-qubit} case of
2.00167 for $\Omega^{HS}_{4}$. For the {\it qubit-qutrit} 
analysis, $7 \times 10^9$ 
points in 35- and 34-dimensional hypercubes were utilized. An estimate of 2.0279 was obtained for $\Omega^{HS}_{6}$
\cite[sec.~VI C 1]{slaterPRA}. This was a (``pooled'') average of two estimates, based on two {\it 
inequivalent} ways of computing the PPT. 
When we generated the partial transpose by transposing in place the four 
$3 \times 3$ blocks of the corresponding $6 \times 6$ density matrices, our estimate was 1.99954, while when we transposed in place the nine $2 \times 2$ 
blocks, we obtained 2.05803 \cite[last paragraph]{slaterPRA}.
(Unfortunately, we lack any particular explanation for why one estimate should be so close to 2, while the other is 
relatively distant. It would appear that
either some form of 
numerical instability is at work, or that the two types of
partial transposition --- surprisingly/puzzlingly --- give rise to different 
(but close) 
Hilbert-Schmidt ratios. As illustrated in \cite[sec. IV]{slaterPRA}, 
a $6 \times 6$
density matrix can have a PPT under one such form of partial transposition,
but not the other.)

Additionally, in \cite[sec. VI C 3]{slaterPRA}, we had  reported
the early stages of a 
{\it qutrit-qutrit} ($N=M=3$) analysis. This was based on
a similar procedural 
scheme with, at that stage,
$126 \times 10^6$ Tezuka-Faure 80- and 79-dimensional 
points having been already generated. (To perform the required 79-dimensional
numerical integration, we merely extracted, an essentially arbitrary 
subset from
the 80-dimensional set, rather than going to the considerably greater 
computational expense of generating a totally new set of Tezuka-Faure 
points {\it ab initio}.)
We reported the achieving, at a late stage of the analysis, 
of a cross-ratio
of 1.89125, which we thought was encouragingly close to 2, for such an
extraordinarily-challenging high-dimensional
numerical integration problem. But we also 
had commented there 
that this ratio seemed subject to large fluctuations
\cite[Fig.~14]{slaterPRA}, and in fact had further  indicated
 that at our last point generated,
the ratio had sunk to approximately 0.15.

We have since continued this same numerical integration process 
from $126 \times 10^6$
to $500 \times 10^6$ (systematically-generated ``low discrepancy'') 
points, and now 
look at it in more detail than in \cite{slaterPRA}. 
(In our analysis, we also include
companion results based on the {\it cross-norm} criterion 
\cite{rudolph} --- rather than
partial transposition --- which added 
substantially to our computational burden. We will only 
briefly discuss these results, as no particular theory/conjectures have
been advanced as pertains to them. The cross-norm was {\it not} included
in our earlier qubit-qubit and qutrit-qutrit analyses \cite{slaterPRA}, so
we have little basis for developing possible relevant hypotheses.)

We recorded the results of the 80-dimensional qutrit-qutrit 
numerical integration, at intermediate 
stages, for 250 intervals, each based on
$2 \times 10^6$ points. 
For the 54-{\it th} to 61-{\it st} such 
intervals, our {\it cumulative}
 estimates of the Hilbert-Schmidt cross-ratio were
\begin{equation} \label{1seq}
\{1.85599, 1.85619, 1.85765, 1.85915, 1.85941, 1.88103, 1.89082, 1.89125\}.
\end{equation}
(Note the monotonic {\it increase} in the direction of the conjectured exact 
value of 2, as more points are
sampled.)

\begin{figure}
\includegraphics{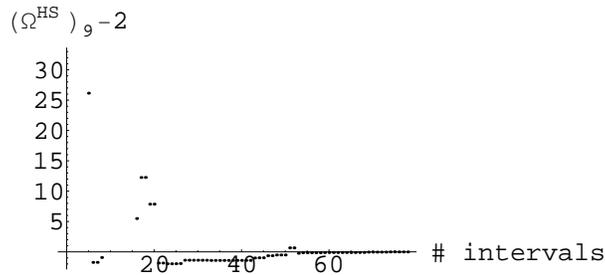}
\caption{\label{fig:editedsequence}Cumulative estimates of $(\Omega^{HS}_{9} -2)$ based upon the first 78 intervals, after the 
discardings of two intervals (marked by large fluctuations) 
from the original 250-long sequence}
\end{figure}

Then, at the 62-{\it nd} interval, the estimate 
sharply plummeted to 0.208052.
Now, we found that 
if we discard/ignore this interval (large fluctuation), and,
then, in the resulting
249-member sequence  of revised 
estimates, discard the 67-{\it th} interval, where the estimate
drops similarly 
from 1.89181 to 0.0198977, we get in the new 248-member sequence of 
re-revised 
cumulative estimates,  a 17-long {\it additional} sequence,
\begin{equation} \label{2seq}
\{1.89083, 1.89098, 1.89101, 1.89056, 1.89181, 1.89892, 1.9031, 1.95864, 
\end{equation}
\begin{displaymath}
1.96107, 1.95866, 1.95938, 1.95924, 1.98872, 1.98913, 1.98842, 1.98853, \
1.98835\}.
\end{displaymath}
which immediately follows upon the 8-long sequence of near-2 estimates
(\ref{1seq}), 
which was extracted from the original unedited 250-member sequence.
So, we have --- after only the two discardings --- a 
consecutive/uninterrupted  sequence of length 25 (= 8 + 17), all 
lying in the range [1.85,2].
Immediately following the last (25-{\it th}) 
member of this sequence (1.98835), 
the estimate first jumps to 2.09132 and then precipitously falls to
0.0786266. In Fig.~\ref{fig:editedsequence}, we plot the 
cumulative estimates of
$(\Omega^{HS}_{9} -2)$ for the edited sequence (containing 78 intervals, each based
on $2 \times 10^6$ points) that ends just before this jump to 2.09132.

We could continue to similarly study the remainder of the  248-member 
sequence, discarding
large fluctuations as they occur (of course, there is the difficult 
question of 
precisely defining them). But, at this stage, it would seem that we have 
already --- coupled with our earlier qubit-qubit and qubit-qutrit 
results \cite{slaterPRA} --- made a 
{\it prima facie} case for the plausibility of the \.Zyczkowski
conjecture stated above, which calls for its further investigation.

It would have been  particularly appealing
if a rational scheme had been 
found for editing the data, with the result
that the {\it final} member of the sequence was near to 2. (As it stands, 
the last/250-{\it th} member of the [unedited] sequence did give 
us an estimate of
18.764 for the Hilbert-Schmidt ratio.) 
Of course, one might continue to 
add points, {\it via} the quasi-Monte Carlo procedure,  to the $500 \times
10^6$ already generated, in the (ensured \cite{giray2}) 
expectation that convergence would 
{\it eventually} occur.

For the same (75-{th}) interval for  which the Hilbert-Schmidt cross-ratio is 
closest to 2 
(that is 1.98913), we obtained for the corresponding cross-ratio ($\Omega^{Bures}_{9}$)
based,  alternatively,
on the Bures metric, an estimate of 
0.260831; for the ``arithmetic-average'' 
metric, 0.236706; for the Wigner-Yanase metric, 0.172202; for the
GKS/quasi-Bures metric, 0.21092; and for (the numerically rather unstable)
Kubo-Mori metric, 0.0116725. 
(Such [{\it monotone} metric \cite{petzsudar}] 
ratios were all close to 2 
in the qubit-qutrit analysis, and approximately 1.8 in the qubit-qubit study 
\cite{slaterPRA}.) 

For the ratios
 based on the cross-norm rather than the PPT criterion,
for which there are currently no conjectures,
the analogous estimates at the same (1.98913) 
interval were --- for the monotone
metrics --- 0.151657 [Bures], 16.718 
[Kubo-Mori], 0.579336 [arithmetic-average], 0.771286 [Wigner-Yanase], 0.287065 
[GKS/quasi-Bures] and for the {\it non}-monotone Hilbert-Schmdit metric 
\cite{ozawa}, 
0.226471. (The corresponding estimates at the termination of the entire unedited 250-long 
sequence were, 
in the same order, 
0.101299, 16.4129, 0.489673, 0.645947, 0.207893 and 0.223288.) 
The analysis also indicated that the cross-norm criterion is 
{\it much} weaker than the 
PPT in distinguishing states that could possibly be separable.
(In sampling points over the 79- and 80-dimensional unit hypercubes 
that served as the domains of integration,
roughly twenty-five times more points [$9 \times 9$ 
density matrices] 
passed the cross-norm test than the PPT test. 
{\it None} at all that passed the cross-norm test failed the PPT one.)

The critical 
reader may have observed that we have not subjected any of the results
above to {\it statistical} testing ---- 
the use of confidence intervals {\it etc.}
The Tezuka-Faure procedure \cite{faure,giray1} 
we have employed is highly efficient in finding
well-distributed (low discrepancy) points, but does not lend itself in any natural fashion
to statistical testing (there are variants, though, that do 
\cite{hong,tuffin,giray2}), 
such as with the much less efficient (random number) Monte Carlo methods.
(The convergence rate of quasi-Monte Carlo is of order 
$n^{-1+p \{\log n \}^{-1/2}}$, where $n$ is the dimension of the problem 
and $p$ is a positive 
number. This is a worst case result. Compared to the expected rate $n^{-1/2}$
of Monte Carlo, it shows the superiority of quasi-Monte Carlo 
\cite{papa,wang}.)

This lack of statistical testability was, to some extent, compensated for
in our earlier 
qubit-qubit and qubit-qutrit analyses \cite{slaterPRA}, 
by the {\it availability} of
exact formulas \cite{zs1,hans1} 
for the Hilbert-Schmidt and Bures volumes and hyperareas
of the $N \times N$ density matrices, against which we could compare
our numerical results, and thus assess their accuracy. 
They  easily came within 1\% of the true formulas in those two instances.

It would be of interest, as well,  to study (PPT/separability) probability 
ratios for $NM \times NM$ density matrices of ranks {\it less} than 
$NM$ and $(NM-1)$ \cite[sec.~VI C 4]{slaterPRA} (cf. \cite{lockhart,albeverio,pawel}). We are also, presently, investigating issues of a similar nature to those analyzed above in the ($NM=8$) case of 
{\it three}-qubit states \cite{dur,rajagopal,guhne}.

If the basic conjectured relation (\ref{firstdefinition}) holds, then one can deduce, using the {\it known} HS total-area-to-total-volume 
ratio \cite[eq. (6.5)]{zs1}, that the ratio of the $(N^2 M^2-1)$-volume of PPT states to the $(N^2 M^2 -2)$-hyperarea
of PPT states must, in general,  be equal to 
 $2/\Big(\sqrt{NM (NM-1)} (N^2 M^2-1) \Big)$. In \cite[sec. VI D  2]{slaterPRA}) this relation, coupled with our numerical integration results, 
 was used to hypothesize certain simple {\it exact} (well-fitting) 
values for 
the HS volumes and hyperareas {\it individually} of the separable
qubit-qubit and qubit-qutrit states. Then, one 
immediately has implied  formulas
for the probabilities $P^{[HS,rank-NM]}_{NM}$ and $P^{[HS,rank-(NM-1)]}_{NM}$.

The quasi-Monte Carlo (Tezuka-Faure) numerical integrations conducted here
and in \cite{slaterPRA} over the domains (unit hypercubes) of the 
$NM \times NM$ density matrices, were all
greatly facilitated by the use of the corresponding {\it Euler angle}
parameterizations \cite{toddtilma}.
\begin{acknowledgments}
I wish to express gratitude to the Kavli Institute for Theoretical
Physics (KITP)
for computational support in this research. This project 
was undertaken at the 
suggestion of K. \.Zyczkowski.
I would like to thank him for his kindness and encouragement 
over the last several years.

\end{acknowledgments}

\bibliography{Conjecture}% Produces the bibliography via BibTeX.

\end{document}